
\magnification=\magstephalf
\documentstyle{amsppt}
\loadbold
\overfullrule=0pt
\def\BC{\Bbb C}

\def\BZ{{\Bbb Z}}
\def\mathD{{\Cal D}}
\let\CD\mathD
\def\mathP{{\Cal P}}
\def\gl{{gl}}
\def\resp{\text{resp.}\, }
\def\fb{{\frak b}}
\def\fg{{\frak g}}
\def\fn{{\frak n}}
\def\CU{{\Cal U}}

\def\Aut{\operatorname{Aut}}
\def\Der{\operatorname{Der}}
\def\End{\operatorname{End}}
\def\Hom{\operatorname{Hom}}
\def\Res{\operatorname{Res}}
\def\ch{\operatorname{ch}}
\def\diag{\operatorname{diag}}
\def\tr{\operatorname{tr}}

\def\psup#1{{}^{#1}\!}

\topmatter

\title
Representation theory of the vertex algebra $\boldsymbol
W_{\bold 1 \boldkey + \boldsymbol \infty}$
\endtitle

\author
Victor Kac and Andrey Radul
\endauthor

\address
Department of Mathematics, MIT, Cambridge, MA 02139,
USA.
\endaddress

\email
kac\@math.mit.edu
\endemail

\thanks
Supported in part by NSF grant DMS--9103792.
\endthanks

\address
Department of Mathematics, Howard University,
Washington, D.C., USA.
\endaddress

\email
aor\@math.umd.edu
\endemail

\endtopmatter

\document

\head Abstract \endhead

In our paper~\cite{KR} we began a systematic study of
representations of the universal central extension $\widehat{\Cal
D}\/$ of the Lie algebra of differential operators on the circle.
This study was continued in the paper~\cite{FKRW} in the
framework of vertex algebra theory.  It was shown that the
associated to $\widehat {\Cal D}\/$ simple vertex algebra $W_{1+
\infty, N}\/$ with positive integral central charge $N\/$ is
isomorphic to the classical vertex algebra $W (gl_N)$, which led
to a classification of modules over $W_{1 + \infty, N}$.  In the
present paper we study the remaining non-trivial case, that of a
negative central charge $-N$.  The basic tool is the
decomposition of $N\/$ pairs of free charged bosons with respect
to $gl_N\/$ and the commuting with $gl_N\/$ Lie algebra of
infinite matrices $\widehat{gl}$.

\head Introduction \endhead

In this paper we study representations of a central extension
$\hat\mathD = \mathD \oplus \BC C\/$ of the Lie algebra
$\mathD\/$ of differential operators on the circle.  This central
extension appeared first in \cite{KP} and its uniqueness was
subsequently established in \cite{F} and in \cite{Li}.

In our paper \cite{KR} we began a systematic study of
representations of the Lie algebra $\hat\mathD$.  In particular,
we classified the irreducible ``quasi-finite'' highest weight
representations and constructed them in terms of irreducible
highest weight representations of the central extension
$\widehat\gl\/$ of the Lie algebra of infinite matrices.

This study was continued in \cite{FKRW} in the framework of
vertex algebra theory.  The advantage of such an approach is
twofold.  From the mathematical point of view it picks out the
most interesting representations and equips them with a rich
additional structure.  From the physics point of view it provides
building blocks of two-dimensional conformal field theories.  We
just mention here applications to integrable systems \cite{ASM},
to $W\/$-gravity \cite{BMP}, and to the quantum Hall effect
\cite{CTZ1,CTZ2}.

In more detail, let us consider the subalgebra $\mathP\/$ of
$\mathD\/$ consisting of the differential operators that can be
extended in the interior of the circle.  The cocycle of our
central extension (see \thetag{6.2}) restricts to a zero cocycle
on $\mathP$, hence $\mathP\/$ is a subalgebra of $\hat\mathD$.
Given $c \in \BC$, we denote by $M_c\/$ the representation of
$\hat\mathD\/$ induced from the 1-dimensional representation of
the subalgebra $\mathP \oplus \BC C\/$ defined by $\mathP \mapsto
0$, $C \mapsto c$.  It was shown in \cite{FKRW} that $M_c\/$
carries a canonical vertex algebra structure.  The
$\hat\mathD\/$-module $M_c\/$ has a unique irreducible quotient
which carries the induced simple vertex algebra structure.
Following physicists, we denote this simple vertex algebra by
$W_{1 + \infty, c}$.

The highest weight representations of the vertex algebra $M_c\/$
are in a canonical 1--1 correspondence with that of the Lie
algebra $\hat\mathD$.  As usual, the situation is much more
interesting when we pass to the simple vertex algebra $W_{1 +
  \infty, c}$.  The cocycle \thetag{6.2} is normalized in such a
way that $M_c\/$ is an irreducible $\hat\mathD\/$-module ({\it
  i.e.}, $M_c = W_{1+\infty, c}$) iff $c \notin \BZ$.

Thus, we are led to the problem of classification of irreducible
highest weight representations of the vertex algebra $W_{1 +
  \infty, c}\/$ with $c \in \BZ$.  This is a highly non-trivial
problem since the field corresponding to every singular vector of
$M_c\/$ must vanish in a representation of $W_{1 + \infty, c}\/$
which gives rise to an infinite set of equations.

The problem has been solved in \cite{FKRW} in the case of a
positive integral $c = N\/$ by the use of an explicit isomorphism
of $W_{1 + \infty, N}\/$ and the classical $W\/$-algebra $W
(\gl_N)$.  This approach cannot be used for $c\/$ negative since
probably $W_{1 + \infty, - N}\/$ is a new structure, which is not
isomorphic to any classical vertex algebra.

Another important result of \cite{FKRW} is an explicit
construction of a ``model'' of ``integral'' representations of
$W_{1 + \infty, N}\/$ by decomposing the vertex algebra of $N\/$
charged free fermions with respect to the commuting pair of Lie
algebras $\gl_N\/$ and~$\widehat\gl$.

One of the main results of this paper is an analogous
decomposition of the vertex algebra of $N\/$ charged free bosons
with respect to a commuting pair $\gl_N\/$ and $\widehat\gl$,
which produces a large class of irreducible modules of the vertex
algebra $W_{1 + \infty, -N}\/$ (Theorems~3.1 and 6.1).  We
conjecture that all irreducible modules of $W_{1 + \infty, -N}\/$
can be obtained from these by restricting to $\hat\mathD\/$ and
taking their certain tensor products (Conjecture~6.1).  This
explicit construction allows one to also derive explicit
character formulas (formula \thetag{3.7}).  Partial results in
this direction were previously obtained in \cite{Mat} and
\cite{AFMO}.

Our basic tool is the suitably modified theory of dual pairs of
Howe \cite{H1}, \cite{H2}.  Namely one has the following general
irreducibility theorem (Theorem~1.1): if a Lie algebra $\frak
g\/$ acts completely reducibly on an associative algebra $A\/$
and if $V\/$ is an irreducible $A\/$-module with an equivariant
action of $\frak g$ such that $V\/$ is a direct sum of at most
countable number of irreducible finite-dimensional modules, then
the pair $\frak g$ and $A^{\frak g} = \left\{ a \in A \mid \frak
g a = 0 \right\}$ acts irreducibly on each isotypic component of
$\frak g$ in~$V$.

This result allows us not only to decompose both free charged
fermions and bosons with respect to the pair $\gl_N\/$ and
$\widehat\gl$, but also to interpret the vertex algebra $W_{1 +
  \infty, -N}\/$ as a subalgebra of the vertex algebra of $N\/$
free charged bosons killed by $\gl_N\/$ (formula~\thetag{4.4}).
In the same way, one identifies $W_{1 + \infty, N}\/$ with a
subalgebra of the vertex algebvra of $N\/$ free charged fermions
killed by $\gl_N$, a result previously obtained in \cite{FKRW}.
This result is interpreted in that paper as an isomorphism $W_{1
  + \infty, N} \simeq W (\gl_N)$ due to the connection to affine
Kac-Moody algebras.  It seems, however, that there is no such
interpretation in the case of negative central charge.

It is interesting to note that while the decomposition of $N\/$
free charged fermions produces all unitary (with respect to the
compact involution) irreducible highest weight representations of
$\widehat{gl}\/$ with central charge $N$, the decomposition of
$N\/$ free charged bosons produces a very interesting class of
irreducible highest weight representations of $\widehat{gl}\/$
with central charge $-N$.  This allows us to compute their
characters (formula \thetag{3.7}).  We also show that the
subcategory of the category $O\/$ of representations all of whose
irreducible subquotients are members of this class is a
semisimple category (Theorem~4.1).  (Of course, in the first case
a similar result goes back to H.~Weyl.)  Provided that
Conjecture~6.1 is valid, this implies semisimplicity of the
category of positive energy $W_{1 + \infty, -N}\/$-modules.

The paper is organized as follows.  In Section~1 we give a proof
of the general irreducibility theorem (Theorem~1.1).  In
Section~2 we use Theorem~1.1 and classical invariant theory (as
in \cite{H2}) to prove irreducibility of each isotypic component
of $\gl_N\/$ in the metaplectic representation of the infinite
Weyl algebra $W_N\/$ ($N\/$ free charged bosons in physics
terminology) with respect to the commuting pair $\gl_N\/$
and~$\widehat\gl\/$ (Theorem~2.1).  By a somewhat lengthy
combinatorial argument, we derive an explicit highest weight
correspondence in Section~3 (Theorem~3.1) and a character formula
for the above-mentioned highest weight
representations of $\widehat\gl\/$~(formula~\thetag{3.7}).  In
Section~4 we prove the complete reducibility Theorem~4.1.
Section~5 is a brief digression on vertex algebras and their
twisted modules which we conclude by a construction of twisted
modules over $N\/$ free charged bosons which become untwisted
modules with respect to $W_{1 + \infty, -N}$.  In Section~6 we
construct a large family of representations of the vertex algebra
$W_{1 + \infty, -N}\/$ using the above mentioned modules
(Theorem~6.2) and conjecture that these are all its irreducible
representations (Conjecture 6.1).  In Section~7 we apply similar
methods to $N\/$ free charged fermions to recover most of the
results of~\cite{FKRW}.

\head 1. Representations of associative $\frak g$-algebras
\endhead

Let $A\/$ be an associative algebra over $\BC$ and let $\Der
A\/$ denote the Lie algebra of derivations of $A$.  Let
$\fg$ be a Lie algebra over $\BC\/$ and let $\varphi : \fg \to
\Der A\/$ be a Lie algebra homomorphism.  The triple $(A, \fg,
\varphi)$ is called an {\it associative $\fg$-algebra}.

An $A\/$-module $V\/$ is called a $(\fg, A)$-{\it module\/}
if $V\/$ is given a structure of a $\fg$-module such that the
$A\/$-module structure is equivariant, i.e.
$$
  g (a v) = (\varphi (g) a ) v + a (gv), \quad
  \text{$g \in \fg$, $a \in A$, $v \in V$}.
$$

Let $A^{\fg} = \{ a \in A \mid \varphi (\fg) a = 0$ for all $g
\in \fg \}$ be the centralizer of the action of $\fg\/$ on $A$.
Given a $\fg\/$-submodule $U\/$ of $V\/$ and $a \in A^\fg$, the
map $U \to a U$, given by $u \mapsto a u$, is clearly a
$\fg$-module homomorphism.

Given an irreducible $\fg$-module $E$, denote by $V_E\/$
the sum of all $\fg\/$-submodules of $V\/$ isomorphic to $E$.
This is called the $E\/$-{\it isotypic component\/} of the
$\fg$-module $V$.  By the above remark, $V_E\/$ is a
$A^\fg$-submodule of $V$.

Choose a 1-dimensional subspace $f \subset E$.  Then, due to
Schur's lemma, this gives us a choice of a 1-dimensional subspace
in each of the irreducible $\fg$-submodules of $V_E$.  We
denote the sum of all of these 1-dimensional subspaces by
$V^E\/$ (it depends on the choice of $f\/$).  Clearly,
$V^E\/$ is a $A^\fg\/$-submodule of $V_E$, and we have a
(non-canonical) $(\fg, A^\fg)$-module isomorphism:
$$
  V_E \simeq E \otimes V^E.
$$
The Lie algebra $\fg$ (resp.\ associative algebra $A^\fg\/$)
acts on $E \otimes V^E\/$ by $g (e \otimes v) = g e \otimes
v\/$ (resp.\ $a (e \otimes v) = e \otimes a v$).

Thus, if $V\/$ is a $(\fg, A)$-module, which is a semisimple
$\fg$-module, we have the following isomorphism of $(\fg,
A^\fg)$-modules:
$$
  V \simeq \bigoplus_E
  \left(
    E \otimes V^E
  \right)
  \tag{1.1}
$$
where summation is taken over all equivalence classes of
irreducible $\fg$-modules.

We wish to study the situations when each isotypic component
$V_E\/$ is irreducible as a $(\fg, A^\fg)$-module, or
equivalently,  when each $A^\fg$-module $V^E\/$ is
irreducible.

\proclaim{Theorem 1.1}
  Let $A$ be a semisimple $\fg$-module.  Let $V$ be a
  $(\fg, A)$-module such that
  \roster
  \item"(i)" $V$ is an irreducible $A$-module

  \item"(ii)" $V$ is a direct sum of at most countable number of
    finite-dimensional irreducible $\fg$-modules.
  \endroster
  Then each isotypic component $V_E\/$ (where $E$ is
  a finite-dimensional irreducible $\fg$-module) is an
  irreducible $(\fg, A^\fg)$-module; equivalently, each $V^E$
  is an irreducible $A^\fg$-module.
\endproclaim

The proof of the theorem is based on some simple lemmas.

\proclaim{Lemma 1.1 (cf.\ \cite{H2})}
  Each $V^E\/$ is an irreducible $(\End V)^\fg$-module.
  (Recall that $\fg$ acts on $\End V$ by $(g a)v = g (av) -
  a (gv)$, $g \in \fg$, $a \in \End V$, $v \in V$.)
\endproclaim

\demo{Proof}
  Let $x, y \in V^E$.  We have the following decomposition in a
  direct sum of $\fg$-modules:
  $$
    V = (E \otimes x) \oplus (E \otimes y) \oplus \CU.
  $$
  Then $a \in \End V\/$ defined by $(f_i \in E$, $u \in
  \CU)$:
  $$
    a
    \left(
      f_1 \otimes x + f_2 \otimes y + u
    \right) = f_1 \otimes y + f_2 \otimes x + u
  $$
  lies in $(\End V)^\fg$.  This proves the lemma.\qed
\enddemo

\proclaim{Lemma 1.2 (cf.\ \cite{H2})}
  Let $X$ be a finite-dimensional $\fg$-invariant subspace of
  $V$.  Then the $\fg$-module $\Hom_\BC (X, V)$ is
  semisimple and the $\fg$-module map $A \to \Hom_\BC (X,
  V)$ (defined by $a \mapsto \varphi_a$ where $\varphi_a (x)
  = a x$, $x \in X$) commutes with projections on isotypic
  components.
\endproclaim

\demo{Proof}
  Let $U = \Hom_\BC (X, V)$.  Since $\dim X < \infty$, we
  have the $\fg$-module isomorphism $U \simeq X^* \otimes V$,
  hence $U$ is a semisimple $\fg$-module:
  $$
    U = \bigoplus_E
    \left(
      E \otimes U^E
    \right).
  $$
  Since $A\/$ is a semisimple $\fg$-module, we have:
  $$
    A = \bigoplus_E
    \left(
      E \otimes A^E
    \right).
  $$
  Let $A_X = \{ a \in A \mid a X = 0 \}$.  This is a
  $\fg$-submodule of $A$, so that
  $$
    A_X = \bigoplus_E
    \left(
      E \otimes A^E_X
    \right), \quad
    A^E_X \subset A^E,
  $$
  and, clearly the map $A \to U\/$ coincides with the canonical
  map
  $$
    \oplus
    \left(
      E \otimes A^E
    \right) \to \oplus
    \left(
      E \otimes
      \left(
        A^E / A^E_X
      \right)
    \right).
  $$
  This proves the lemma.\qed
\enddemo

\proclaim{Lemma 1.3} (Jacobson's density theorem)
  Let $V$ be at most countable-dimensional irreducible $A$-module.
  Then for any finite-dimensional subspace $X$ of $V$ and any $f
  \in \End_\BC V$ there exists $a \in A$ such that
  $$
    f (x) = a (x) \text{ for all } x \in X.
  $$
\endproclaim

\demo{Proof}
  $\End_A V = \BC\/$ since $V\/$ is irreducible and at most
  countable-dimensional.  Now lemma follows from Jacobson's
  density theorem as stated in \cite{L}.\qed
\enddemo

\demo{Proof of Theorem~1.1}
  Let $x, y \in V^E$.  By Lemma~1.1 there exists $f
  \in (\End V)^\fg$ such that $f (x) = y$.  By
  Lemma~1.3 there exists $a \in A\/$ such that $a\/$
  coincides with $f\/$ on the subspace $U = E \otimes x + E
  \otimes y$.  Let $a_0$ be the projection of $a\/$ on
  $A^\fg$.  By Lemma~1.2, $a_0$ still coincides
  with $f\/$ on $U$.  Indeed, we have for $u \in U$, the
  subscript $E\/$ denoting the projection on $V_E$:
  $$
    \alpha_0 (u) = a (u)_E = f (u)_E = f (u).\eqno\qed
  $$
\enddemo

We do not know whether one can remove the finite-dimensionality
assumption in (ii).  The following proposition shows that one can
for the isotypic component of the trivial $\fg$-module.

\proclaim{Proposition 1.1}
  Let $A$ be an associative $\fg$-algebra, completely
  reducible with respect to $\fg$.  Let $V$ be an $(A,
  \fg)$-module which is irreducible with respect to $A$ and
  completely reducible with respect to $\fg$.  Then $V^\fg :=
  \{ v \in V \mid \fg v = 0\}$ is an irreducible
  $A^\fg$-module.
\endproclaim

\demo{Proof}
  We have the decompositions into isotypic components with
  respect to $\fg$:
  $$
    A = A^\fg \oplus \sum_{E \neq 1} A_E, \quad
    V = V^\fg \oplus \sum_{E \neq 1} V_E,
  $$
  where 1 stands for the 1-dimensional trivial representation of
  $\fg$.  Note that
  $$
    \left(
      \sum_{E \neq 1} A_E
    \right) V^\fg \subset \sum_{E \neq 1} V_E, \quad
    A^\fg V^\fg \subset V^\fg.
  $$
  Let $v \in V^\fg$ be a non-zero vector.  Since $V\/$ is an
  irreducible $A\/$-module, we have:  $A v = V$.  Hence:
  $$
    V = A v = A^\fg v \oplus
    \left(
      \sum_{E \neq 1} A_E
    \right) v \subset A^\fg v \oplus \sum_{E \neq 1} V_E.
  $$
  It follows that $A^\fg v = V^\fg$.\qed
\enddemo

\remark{Remark 1.1}
  Theorem~1.1 and Proposition~1.1 have
  obvious group analogues.  Let $G\/$ be a group and let
  $\varphi : G \to \Aut A\/$ be a group homomorphism.  This is
  called an associative $G\/$-algebra.  An $A\/$-module
  $V\/$ is called a $(G, A)$-module if $V\/$ is given a
  structure of a $G\/$-module such that the $A\/$-module
  structure is $G\/$-equivariant, i.e.
  $$
    g (av) = (\varphi (g) a) (g v), \quad
    \text{$g \in G$, $a \in A$, $v \in V$
      }.
  $$
  Then Theorem~1.1 and Proposition~1.1 hold
  with $\fg$ replaced by $G$.
\endremark

\head 2. Charged free bosons \endhead

Consider $N\/$ pairs of free charged bosonic fields ($i = 1,
\ldots, N\/$):
$$
  \gamma^i (z) = \sum_{m \in \BZ} \gamma^i_m z^{-m - 1}, \quad
  \gamma^{*i} (z) = \sum_{m \in \BZ} \gamma^{*i}_m z^{-m}.
$$
Recall that this is a collection of local even fields with the
OPE ($i, j = 1, \ldots, N\/$):
$$
  \gamma^{*i} (z) \gamma^j (w) \sim \frac{\delta_{ij}}{z - w},
  \quad
  \text{all other OPE } \sim 0,
  \tag{2.1}
$$
with the vacuum vector $| 0 \rangle$ subject to conditions
$$
  \gamma^i_m | 0 \rangle = 0 \text{ for } m \geq 0, \quad
  \gamma^{*i}_m | 0 \rangle = 0 \text{ for } m > 0,.
  \tag{2.2}
$$
In other words, we have a unital associative algebra, usually
called the {\it Weyl algebra\/} and denoted by $W_N$, on
generators $\gamma^i_m$, $\gamma^{*i}_m\/$ ($i = 1, \ldots,
N$; $m \in \BZ$) with the following defining relations:
$$
  \left[
    \gamma^{*i}_m, \gamma^j_n
  \right] = \delta_{ij} \delta_{m, -n} \quad
  \left[
    \gamma^i_m, \gamma^j_n
  \right] = 0 =
  \left[
    \gamma^{*i}_m, \gamma^{*j}_n
  \right].
  \tag{2.3}
$$
This algebra has a unique irreducible representation in a vector
space $M$, called the {\it metaplectic representation}, such
that there exists a non-zero vector $|0 \rangle$ satisfying conditions
\thetag{2.2}.

We let
$$
  \gather
    e^{ij} (z) = : \gamma^i (z) \gamma^{*j} (z) : \equiv
    \sum_{m \in \BZ} e^{ij}_m z^{-m-1} \quad (i, j = 1, \ldots,
    N), \\
    E (z, w) = \sum^N_{p = 1} : \gamma^p (z) \gamma^{*p} (w) :
    \equiv \sum_{i, j \in \BZ} E_{ij} z^{i - 1} w^{-j},
  \endgather
$$
where the normal ordering $: :$ as usual means that the
operators annihilating $| 0 \rangle$ are moved to the right.

Using Wick's formula, it is immediate to see that the operators
$e^{ij}_m\/$ form a representation in $M\/$ of the affine
Kac-Moody algebra $\widehat\gl_N\/$ of level $-1$
\cite{K1}:
$$
  \left[
    e^{ij}_m, e^{st}_n
  \right] = \delta_{j s} e^{i t}_{m + n} - \delta_{i t} e^{s
    j}_{m + n} - m \delta_{m, -n} \delta_{j s} \delta_{i t}.
  \tag{2.4}
$$
In particular, the operators $e_{i j} : = e^{i j}_0$ form a
representation of the general linear Lie algebra $\gl_N$.  We
have also the following commutation relations:
$$
  \left[
    e^{i j}_m, \gamma^k_n
  \right] = \delta_{j k} \gamma^i_{m + n}, \quad
  \left[
    e^{ij}_m, \gamma^{* k}_n
  \right] = - \delta_{ik} \gamma^j_{m + n}.
  \tag{2.5}
$$
Hence the Lie algebra $\widehat \gl_N\/$ acts on the Weyl
algebra $W_N\/$ by derivations via the adjoint representation
$g (a) = [g, a]$.

The following important relation is straightforward:
$$
  \left[
    e_{ij}, E_{mn}
  \right] = 0 \quad \text{for all} \quad
  i, j = 1, \ldots, N;\ m, n \in \BZ.
  \tag{2.6}
$$

Furthermore, the operators $E_{ij}\/$ ($i, j \in \BZ$) form a
representation in $M\/$ of the Lie algebra $\widehat\gl\/$ with
central charge $-N$:
$$
  \left[
    E_{ij}, E_{st}
  \right] = \delta_{j s} E_{i t} - \delta_{i t} E_{s j} - N \Phi
  \left(
    E_{ij}, E_{st}
  \right),
  \tag{2.7}
$$
where the 2-cocycle $\Phi\/$ is defined by:
$$
  \Phi (A, B) = \tr ([J, A] B), \quad J = \sum_{i \leq 0} E_{i i}.
$$
Recall that $\widehat\gl = \widetilde\gl + \BC K\/$ is a central
extension defined by the cocycle $\Phi\/$ of the Lie algebra
$\widetilde\gl\/$ of all matrices $(a_{ij})_{i, j \in \BZ}$ with
finitely many non-zero diagonals.  We have also the following
commutation relations: %
$$
  \left[
    E_{i j}, \gamma^k_{-m}
  \right] = \delta_{j m} \gamma^k_{-i}, \quad
  \left[
    E_{i j}, \gamma^{* k}_m
  \right]
  = - \delta_{i m} \gamma^{* k}_j.
  \tag{2.8}
$$
These formulas define a representation of $\widehat\gl\/$ on
$W_N\/$ by derivations.

Introduce the following subspaces of the algebra $W_N$:
$$
  \everymath{\displaystyle}
  \spreadmatrixlines{1.5ex}
  \matrix\format\r&\c&\l&\qquad\r&\c&\l\\
    U_m & = & \sum_{i = 1}^N \BC \gamma^i_{-m}, &
    U^*_m & = & \sum^N_{i = 1} \BC \gamma^{*i}_m, \\
    U & = & \sum_{m \in \BZ} U_m, &
    U^* & = & \sum_{m \in \BZ} U^*_m, \\
    U_- & = & \sum_{m \leq 0} U_m, &
    U^*_- & = & \sum_{m \leq 0} U^*_m, \\
    U_+ & = & \sum_{m > 0} U_m, &
    U^*_+ & = & \sum_{m > 0} U^*_m.
  \endmatrix
$$
The following observations are clear by \thetag{2.5}.

\remark{Remark 2.1}
  The subspaces $U_m\/$ (resp.\ $U^*_m\/$) are
  $\gl_N\/$-submodules of $W_N\/$ isomorphic to the standard
  $\gl_N\/$-module $\BC^N\/$ (resp.\ its dual).
\endremark

\remark{Remark 2.2}
  As a $\gl_N\/$-module (with respect to the adjoint
  representation), $W_N\/$ is isomorphic to the symmetric
  algebra over the $\gl_N\/$-module $U + U^*$.
\endremark

Since $\gl_N | 0 \rangle = 0$, we obtain

\remark{Remark 2.3}
  As a $\gl_N\/$-module, $M\/$ is isomorphic to the symmetric
  algebra over the $\gl_N\/$-module $U^*_- + U_+$.
\endremark

\proclaim{Proposition 2.1}
  The centralizer $(W_N)^{{\gl_N}}\/$ of the action of
  $\gl_N\/$ on the algebra $W_N\/$ is an associative
  subalgebra generated by the elements
  $$
    E_{ij} = \sum^N_{p = 1} : \gamma^p_{-i} \gamma^{*p}_{j} :,
  $$
  where $i, j \in \BZ$.
\endproclaim

\demo{Proof}
  Due to Remarks~2.1 and 2.2, the proposition
  follows from the first fundamental theorem of classical
  invariant theory for $GL_N\/$ (which states that the algebra
  of invariant polynomials on the direct sum of any number of
  copies of the standard $GL_N\/$-module $\BC^N\/$ and its
  dual is generated by the obvious invariant polynomials of
  degree two, see {\it e.g.} \cite{VP}).\qed
\enddemo

Due to Remarks~2.1 and 2.3, $M\/$ decomposes in a direct sum of
finite-dimensional irreducible $\gl_N\/$-modules.  Let %
$$
  M = \bigoplus_E M_E
  \tag{2.9}
$$
be the decomposition of $M\/$ in a direct sum of isotypic
components with respect to $\gl_N$.  It is easy to prove now
the first main result of the paper.

\proclaim{Theorem 2.1}
  Each isotypic component $M_E$ in \rom{\thetag{2.9}} is
  irreducible with respect to the Lie algebra $\gl_N \bigoplus
  \widehat\gl$.
\endproclaim

\demo{Proof}
  Due to Remarks~2.1--2.3, all conditions of
  Theorem~1.1 hold for the\linebreak
  $(\gl_N, W_N)$-module $M$.  It follows that each $M_E\/$ is an
  irreducible $\left( \gl_N, W^{\gl_N}_N \right)$-module.  Now
  the theorem follows from Proposition~2.1.\qed
\enddemo

\head 3. The decomposition of $M\/$ with respect to
$\gl_N \bigoplus \widehat\gl$
\endhead

Let
$$
\everymath{\displaystyle}
\matrix
  \gl_- & = \sum_{i, j \leq 0} \BC E_{i j} \subset W^-_N, \\
  \gl_+ & = \sum_{i, j > 0} \BC E_{i j} \subset W^+_N.
\endmatrix
$$
These are Lie subalgebras of the algebras $W^\pm_N\/$ viewed as
Lie algebras with the usual bracket.  They are also subalgebras
of the Lie algebra $\widehat\gl\/$ (the restriction of the
2-cocycle $\Phi\/$ to these subalgebras is trivial).  Of course
$\gl_-$ (resp.\ $\gl_+$) is naturally identified with the Lie
algebra of all matrices $(a_{i j})$ with only finite number of
non-zero entries where $i, j \leq 0$ (resp.\ $> 0$).

Let $\fb_N\/$ (resp.\ $\fb_\pm$) be the Borel subalgebra of all
upper triangular matrices in $\gl_N$ (resp.\ $\gl_\pm$).  Let
$I_N = \{1, \ldots, N\}$, $I_+ = \{ 1, 2, 3, \ldots \}$,
$I_- = \{\ldots, -2, -1, 0\}$.  Given $\lambda = \{ \lambda_i
\}_{i \in I_N (\text{resp.}~ I_\pm)}$ there exists a unique
irreducible module $L_N (\lambda)$ (resp.\ $L_\pm (\lambda)$)
which admits a (unique up to a constant factor) non-zero vector
$v_\lambda\/$ such that $\BC v_\lambda\/$ is invariant with
respect to the Borel subalgebra $\fb_N\/$ (resp.\ $\fb_\pm$) and
$$
  E_{i i} v_\lambda = \lambda_i v_\lambda, \quad
  i \in I_N \text{(resp. $I_\pm$)}.
$$
These modules are called irreducible highest weight modules (with
highest weight $\lambda$).

For example the standard $\gl_N\/$-module $\BC^N\/$ (resp.\
its dual) is the irreducible module with highest weight $(1, 0,
\ldots, 0)$ (resp.\ $(0, \ldots, 0, -1)$).  Tensor products of
copies of the standard $\gl_N\/$-module (resp.\ its dual)
decompose into a direct sum of finite-dimensional irreducible
modules with highest weights from the set
$$
  H^+_N \text{(resp. $H^-_N$)} =
  \left\{
    (\lambda_i)_{i \in I_N} \mid
    \lambda_i \text{(resp. $-\lambda_i$)} \in \BZ_+,\
    \lambda_i \geq \lambda_j \text{ if } i < j
  \right\}.
$$

The standard $\gl_+ (\resp \gl_-)$-module is $\BC^{+ \infty} =
\bigoplus_{j > 0} \BC v_j$ (\resp $\BC^{-\infty} =$\linebreak
$\bigoplus_{j \leq 0} \BC v_j)$ with the usual action $E_{i j}
v_k = \delta_{j k} v_i$.  The $\gl_+$-module $\BC^{+\infty}$
(\resp $\gl_-$-module $\BC^{-\infty}$) is an irreducible module
with highest weight $(1, 0, 0, \ldots)$ (\resp\linebreak
 $(\ldots, 0, 0,
-1)$).  Tensor products of copies of the standard $\gl_+ (\resp
\gl_-)$ decompose into a direct sum of irreducible modules with
highest weights from the set %
$$
  H_\pm =
  \left\{
    (\lambda_i)_{i \in I_\pm} \mid \pm \lambda_i \in \BZ_+, \
    \lambda_i \geq \lambda_j \text{ if } i < j
  \right\}.
$$
We shall identify $H^\pm_N\/$ with a subjset of $H_\pm$ via
the following map:
$$
  \matrix\format\l&\l\\
    (\lambda_1, \ldots, \lambda_N) & \mapsto \\
    \strut
  \endmatrix
  \left\{
    \matrix\format\l&\l\\
      (\lambda_1, \ldots, \lambda_N, 0, 0, \ldots) &
      \text{in $+$ case}, \\
      (\ldots, 0, 0, \lambda_1, \ldots, \lambda_N) &
      \text{in $-$ case}.
    \endmatrix
  \right.
$$

For convenience of notation we let $\BC^{-N} = \left( \BC^N
\right)^*$.

\proclaim{Lemma 3.1}
  The $\gl_N \bigoplus \gl_\pm$-module $S \left( \BC^{\pm N}
    \otimes \BC^{\pm \infty} \right)$ has the following
  decomposition into a direct sum of irreducible modules:
  $$
    S \left(
      \BC^{\pm N}
      \otimes \BC^{\pm \infty}
    \right) = \bigoplus_{\lambda \in H^\pm_N}
    L_N (\lambda) \otimes L_\pm (\lambda).
  $$
\endproclaim

\demo{Proof}
  The proof follows from the well-known Cauchey's formula (see
  {\it e.g.} \cite{M}):
  $$
    \frac{1}{\displaystyle
      \prod^N_{i = 1} \prod^\infty_{j = 1} (1 - x_i y_j)
      } =
    \sum_{\lambda \in H^+_N} \ch L_N (\lambda) \ch L_+ (\lambda),
  $$
  where $\ch L_N (\lambda) = \tr_{L_N (\lambda)} \diag (x_1,
  \ldots, x_N)$ and $\ch L_+ (\lambda) = \tr_{L_+ (\lambda)}
  \diag (y_1, y_2, \ldots)$ are characters.\qed
\enddemo

Let
$$
  \fg = \gl_N \bigoplus \gl_- \bigoplus \gl_+
$$
be a direct sum of Lie algebras.  Commutation relations
\thetag{2.5} and \thetag{2.8} imply the following

\remark{Remark 3.1}
  We have:
  $$
  \spreadmatrixlines{1.5ex}
    \matrix\format\r&~\c~&\l&~\c&~\c\\
      \left[
        \gl_-, U^*_-
      \right] & \subset & U^*_-,
      \left[
        \gl_-, U_+
      \right] & = & 0, \\
      \left[
        \gl_+, U_+
      \right] & \subset & U_+,
      \left[
        \gl_+, U^*_-
      \right] & = & 0
    \endmatrix
  $$
  Thus, both $U^*_-$ and $U_+$ are $\fg$-modules.
  The $\fg\/$-module $U_+ (\resp U^*_-)$ is isomorphic to
  the $\gl_N \bigoplus \gl_+ (\resp \gl_N \bigoplus
  \gl_-)$-module $\BC^N \otimes \BC^{+\infty} (\resp \BC^{N*}
  \otimes \BC^{-\infty})$ with the trivial action of $\gl_-
  (\resp \gl_+)$.
\endremark

Since $\gl_\pm | 0 \rangle = 0$, we obtain, using also
Remark~2.3:

\remark{Remark 3.2}
  As a $\fg$-module, $M$ is isomorphic to the symmetric
  algebra over\linebreak
  $U^*_- \bigoplus U_+$.
\endremark

It is easy now to decompose the metaplectic representation
$M\/$ with respect to $\fg$.  First, note that, by
Remark~3.2 we have the following isomorphism of
$\fg$-modules
$$
  M \simeq S
  \left(
    U^*_-
  \right) \otimes S
  \left(
    U_+
  \right).
  \tag{3.1}
$$
The decomposition of each of the factors is given by
Lemma~3.1, using Remark~3.1.  In order to
state the result, introduce the following notation.  Given
finite-\linebreak
dimensional irreducible $\gl_N\/$-modules $L_N (\lambda)$ and
$L_N (\mu)$, write their tensor product decomposition:
$$
  L_N (\lambda) \otimes L_N (\mu) = \bigoplus_\nu c_{\lambda,
      \mu}^\nu L_N (\nu), \quad
  c_{\lambda, \mu}^\nu \in \BZ_+.
  \tag{3.2}
$$
Note that if $\lambda \in H^-_N$, $\mu \in H^+_N$, then $\nu
\in H_N$, where
$$
  H_N =
  \left\{
    (\lambda_i)_{i \in I_N} \mid
    \lambda_i \in \BZ, \quad
    \lambda_i \geq \lambda_j \text{ if } i < j
  \right\}.
$$
Thus, we arrive at the following result.

\proclaim{Proposition 3.1}
  The following is a decomposition of $M$ as a $\fg$-module:
  $$
    M = \bigoplus_{
      \Sb
      \lambda \in H^-_N \\
      \mu \in H^+_N \\
      \nu \in H_N
      \endSb
      }
    c_{\lambda, \mu}^\nu L_N (\nu) \otimes L_- (\lambda)
  \otimes L_+ (\mu).
  \eqno\qed
  $$
\endproclaim

Choose in each $\gl_N\/$-module $L_N (\nu)$, $\nu \in
H_N$, the 1-dimensional subspace $\BC v_\nu$.  Then, the
subspace
$$
  M^\nu :=
  \left\{
    m \in M \mid
    \fb_N m \subset \BC m, \quad
    e_{i i} m = \nu_i m
  \right\}
$$
is a $\hat\gl\/$-submodule, which is irreducible due to
Theorem~2.1.  The decomposition \thetag{1.1} becomes:
$$
  M = \bigoplus_{\nu \in H_N} L_N (\nu) \otimes M^\nu
  \tag{3.3}
$$
as $\gl_N \bigoplus \widehat\gl\/$-modules.  Due to
Proposition~3.1, as a $\gl_- \bigoplus \gl_+$-module, $M^\nu\/$
($\nu \in H_N\/$) decomposes as follows:
$$
  M^\nu = \bigoplus_{
    \Sb\lambda \in H^-_N \\
    \mu \in H^+_N
    \endSb
    }
  c_{\lambda, \mu}^\nu L_- (\lambda) \otimes L_+ (\mu).
  \tag{3.4}
$$

Denote by $\fb\/$ the subalgebra of $\widehat\gl\/$ of upper
triangular matrices plus $\BC K$, where $K\/$ is a central
element acting on $M\/$ (and each $M^\nu$) as $-N I$, and let
$\fn$ be the subalgebra of $\fb\/$ of matrices with 0's on the
diagonal.  It is clear that $\fn$ acts locally nilpotently on
$M$.  Hence each $\widehat\gl\/$-module $M^\nu\/$ is an
irreducible highest weight module $L (\Lambda (\nu), -N)$.
Recall that the irreducible highest weight
$\widehat\gl\/$-module $L (\Lambda, c)$, where $\Lambda =
(\Lambda_j)_{j \in \BZ}$ and $c \in \BC$, is defined by the
properties that there exists a unique up to a constant multiple
vector $v \in L (\Lambda, c)\/$ such that $\fn (v) = 0$ and
$$
  E_{i i} v = \Lambda_i v, \quad
  i \in \BZ; \quad K = c I.
$$

We let $\Lambda (\nu) = \left( \Lambda (\nu)_j \right)_{j \in
  \BZ}$.  It remains to calculate $\Lambda (\nu)$ for each
$\nu \in H_N$.

We claim that
$$
  \Lambda (\nu) =
  \left(
    \ldots, 0, \nu_{p + 1}, \ldots, \nu_N;\ \nu_1, \nu_2, \ldots,
    \nu_p, 0, \ldots
  \right),
  \tag{3.5}
$$
where $p = 0$ if all $\nu_i < 0$, $p = N\/$ if all $\nu_i
> 0$, and $1 \leq p < N\/$ is such that $\nu_p \geq 0 \geq
\nu_{p + 1}$ otherwise.  We put here semicolon between the 0th
and the first slots (and comma in all other places).

The proof of \thetag{3.5} will follow from a sequence of
combinatorial lemmas.

Let us define $S\/$ to be the set of pairs of $N\/$-tuples
$(\lambda; \mu)$ such that $\lambda_i, \mu_j \in \BZ$, where
$\lambda_i \leq 0$, $\mu_j \geq 0$ for all $i, j$.

Let us define operations $R^a_{i j} : (\lambda; \mu) \to
(\lambda'; \mu')$, $a = 1, 2, 3$, $1 \leq i, j \leq N\/$
provided that $(\lambda'; \mu') \in S$:
$$
  \spreadmatrixlines{1.5ex}
  \matrix\format\l&\quad\r&~\c&~\l&\quad\r&~\c&~\l&\quad\l\\
    R^1_{i j}: & \lambda'_i & = & \lambda_i - 1, &
    \lambda'_j & = & \lambda_j + 1 & \text{for } i > j, \\
    R^2_{i j}: & \mu'_i & = & \mu_i - 1, &
    \mu'_j & = & \mu_j + 1 & \text{for } i > j, \\
    R^3_{i j}: & \lambda'_i & = & \lambda_i + 1, &
    \mu'_j & = & \mu_j - 1 & \text{for any } i, j,
  \endmatrix
$$
all other entries of $(\lambda; \mu)$ do not change.

We will say that $(\lambda''; \mu'') \geq (\lambda; \mu)$ if
one may obtain $(\lambda''; \mu'')$ from $(\lambda; \mu)$ by
a sequence of operations $R^a_{i j}$.

We define a subset $S_\nu\/$ in $S\/$ by two conditions:
(a)~$\lambda, \mu\/$ are dominant $\gl_N\/$ weights,
(b)~$L_N (\lambda) \otimes L_N (\mu) \supset L_N (\nu)$.

\remark{Remark 3.3}
  A map from $S\/$ into $\widehat\gl\/$-weights given by
  $$
    (\lambda; \mu) \mapsto
    (\ldots, 0, 0, \lambda_1, \ldots, \lambda_N;\ \mu_1, \ldots,
    \mu_N, 0, 0, \ldots)
  $$
  sends $S_\nu\/$ into set of $\gl_- \bigoplus
  \gl_+$-highest weights of the $\widehat\gl\/$-module $L
  (\Lambda (\nu))$ and the ordering on $S\/$ into the standard
  ordering of $\widehat\gl\/$-weights.

  We will describe condition (b) in the definition of $S_\nu\/$ by
  the table
  $$
  \vbox{
    \offinterlineskip
    \halign{
      #\hfil\quad &\strut\vrule\quad#\hfil\cr
      $\mu_1 \dots \mu_p$ & $\mu_{p + 1} \dots \mu_N $\cr
      $\lambda_1 \dots \lambda_p$ & $\lambda_{p + 1} \dots
      \lambda_N$ \cr
      \noalign{\hrule}\cr
      $\nu_1 \dots \nu_p$ & $\nu_{p + 1} \dots \nu_N$ \cr
      \noalign{\hrule} \cr
      $q_1 \dots q_p$ & $q_{p + 1} \dots q_N$ \cr
      }
    }
  $$
  Here $(q_1, \ldots, q_N)$ is a sum of negative $\gl_N\/$
  roots such that $\mu_i + \lambda_i + q_i = \nu_i$.
\endremark

Let $Q^i = \sum_{j \leq i} q_j$.

\proclaim{Lemma 3.2}
  $(q_1, \ldots, q_N)$ is a sum of negative roots if and only
  if
  $$
    Q^i \leq 0; \quad
    Q^N = 0
    \tag{3.6}
  $$
\endproclaim

\demo{Proof}
  Let $q_a\/$ be the negative number with maximal $a$,
  $q_b\/$ the positive number with minimal $b > a$.  If we
  add to $(q_1, \ldots, q_N)$ the (positive) root of $E_{a, b}$
  the condition \thetag{3.6} remains valid and $\sum_i
  |q_i|$ decreases by 2.
\enddemo

\proclaim{Lemma 3.3}
  Let $(\lambda; \mu) \in S_\nu$.  Then
  \roster
  \item"(a)" $\mu_i + q_i \geq 0$,

  \item"(b)" $\lambda_i + q_i \leq 0$,

  \item"(c)" $\mu_j \geq \nu_j$,

  \item"(d)" $\lambda_i \leq \nu_i$.
  \endroster
\endproclaim

\demo{Proof} Since $\lambda_i + \mu_i + q_i = \nu_i$, then (a)
implies (d) and (b) implies (c).  To prove (a) we will use the
description of the decomposition of the tensor product of two
$\gl_N\/$ modules with dominant highest weights based on Weyl
determinants (see \cite{Z}).  In particular this description
implies the following statement.  If $L_N (\lambda) \otimes L_N
(\mu) \supset L_N (\nu)$ and $\mu_i \geq 0$ for all $i$, then one
can find the weight $(\nu_1, \ldots, \nu_N)$ among the
$N\/$-tuples obtained from $(\lambda_1, \ldots, \lambda_N)$ by a
sequence of operations ($m \in \BZ_+$):
  $$
    \Gamma_m (\lambda_1, \ldots, \lambda_N) =
    \sideset \and' \to\sum_{
      \Sb
      \alpha_k \geq 0 \\
      \alpha_1 + \cdots + \alpha_N = m
      \endSb
      }
    (\lambda_1 + \alpha_1, \ldots, \lambda_N + \alpha_N)
  $$
  where one drops $(\lambda_1 + \alpha_1, \ldots, \lambda_N +
  \alpha_N)$ from the sum if at least one of the weights
  $(\lambda_1, \ldots, \lambda_i + \alpha_i, \ldots, \lambda_N)$
  is not dominant.  Hence $\mu_i + q_i \geq 0$.  Since $L_N
  (\lambda)^* \otimes L_N (\lambda)^* \supset L_N (\nu)^*$ the
  same reasoning proves (b).  \enddemo

\proclaim{Lemma 3.4}
Let $r \in [0, p]$, $s \in [p + 1, N]$.  Applying a sequence of
operations $R^1_{s, r}$ and $R^2_{s, r}$ one may get from
$(\lambda; \mu) \in S_\nu$ a sequence $(\lambda'', \mu'') \in S$
such that
  \roster
  \item"(a)" $\mu''_s = 0$; $\lambda''_r = 0$,

  \item"(b)" $\mu''_s + q''_s \geq 0$; $\lambda''_r + q''_r \leq
    0$, where $(q''_1, \ldots, q''_N)$ is a sum of negative
    roots,

  \item"(c)" $q''_r \leq 0$; $q''_s \geq 0$.
  \endroster
\endproclaim

\demo{Proof}
  It is clear that (a) and (b) imply (c).  By Lemma 3.3, (c) is
  valid.  Let $\mu_s > 0$ for some $s$.  Let us apply
  $R^2_{s, r}\/$ for some $r$.  Then $\mu'_s = \mu_s - 1$,
  $\mu'_r = \mu_r + 1$; $q'_s = q_s + 1$, $q'_r = q_r - 1$,
  hence we add a negative root to $(q_1, \ldots, q_N)$.
  Moreover, $\mu_i + q_i\/$ do not change for all $i$;
  $\lambda'_r + q'_r = \lambda_r + q_r - 1$, hence $\lambda'_r
  + q'_r\/$ is still nonpositive for all $r$.  Similar
  considerations apply to $R^1_{s, r}$.\qed
\enddemo

\proclaim{Lemma 3.5}
  Let a pair $(\lambda; \mu) \in S$ be described by a table
  $$
  \vbox{
    \offinterlineskip
    \halign{
      #\hfil\quad &\strut\vrule\quad#\hfil\cr
      $\mu_1 \cdots \mu_p$ & $0 \cdots 0$ \cr
      $0 \cdots 0$ & $\lambda_{p + 1} \cdots \lambda_N$ \cr
      \noalign{\hrule} \cr
      $\nu_1 \cdots \nu_p$ & $\nu_{p + 1} \cdots \nu_N$ \cr
      \noalign{\hrule} \cr
      $q_1 \cdots q_p$ & $q_{p + 1} \cdots q_N$ \cr
      }
    }
  $$
  such that
  \roster
  \item"(a)" $\nu_1 \geq \cdots \geq \nu_p \geq 0 \geq \nu_{p
      + 1} \geq \cdots \geq \nu_N$, $\mu_i + \lambda_i + q_i =
    \nu_i$ where $(q_1, \ldots, q_N)$ is a sum of negative
    roots;

  \item"(b)" $q_r \leq 0$, $q_s \geq 0$.
  \endroster
  Then applying a sequence of operations $R^3_{r, s}$ one may get
  a pair $(\lambda'', \mu'')$ with $q''_i = 0$.
\endproclaim

\demo{Proof}
  If $q_r < 0$ for some $r$, then one may find $s\/$ such
  that $q_s > 0$.  Hence $\mu_r > 0$ and $\lambda_s < 0$,
  and one may apply $R^3_{s, r}$.  It is clear that $Q^i < 0$
  for $i < s$, hence $\{ q'_j \}$ is still a sum of negative
  roots, and $\sum_i |q_i|$ decreases by 2.\qed
\enddemo

\demo{Proof of Theorem~3.1}
It is clear that the sequence on the right of \thetag{3.5} lies
on $S_\nu$.  Lemmas~3.3--3.5 and Remark~3.3 imply that the
corresponding $\widehat\gl\/$-weight is maximal.  This completes
the proof of \thetag{3.5}. \qed
\enddemo

Thus we proved our next main result.

\proclaim{Theorem 3.1}
  With respect to $\gl_N \bigoplus \widehat\gl$ the metaplectic
  representation $M$ decomposes as follows:
  $$
    M = \bigoplus_{\nu \in H_N} L_N (\nu) \otimes L (\Lambda
    (\nu), -N)
    \tag{3.7}
  $$
  where $\Lambda (\nu)$ is defined by \thetag{3.5}.
\endproclaim

For us the most important consequence of this result is

\proclaim{Corollary 3.1}
  $M^{\gl_N} \simeq L (0, -N)$ as $\widehat\gl\/$-modules.
\endproclaim

As another corollary of \thetag{3.7} and \thetag{3.4} , we obtain
a character formula for the irreducible $\widehat\gl\/$-module $L
(\Lambda (\nu), -N)$ with central charge $- N\/$ and highest
weight of the form $\Lambda (\nu)$ given by \thetag{3.5} where
$\nu \in H_N$: %
$$
  \tr_{L (\Lambda (\nu), -N)} \diag
  \left(
    \ldots, y^{-1}_2, y^{-1}_1; x_1, x_2, \ldots
  \right) =
  \sum_{\lambda, \mu \in H^+_N} c_{\lambda^* \mu}^\nu S_\lambda
  (y) S_\mu (x).
  \tag{3.8}
$$
Here for $\lambda = (\lambda_1, \ldots, \lambda_N) \in H^+_N\/$
we let $\lambda^* = (- \lambda_N, \ldots, - \lambda_1)$ ($\in
H^-_N$), $S_\mu (x)$ stands for $\tr_{L_+ (\mu)} \diag
(x_1, x_2, \ldots)$, and $c_{\lambda \mu}^\nu$ are defined by
\thetag{3.2}.  Formula \thetag{3.8} follows from \thetag{3.4} and
the proof of Theorem~3.1.  In particular, for the {\it vacuum\/}
$\widehat\gl\/$-module $L (0, -N)$ with central charge $-N\/$ we
obtain:
$$
  \tr_{L (0, -N)} \diag
  \left(
    \ldots, y^{-1}_2, y^{-1}_1;\ x_1, x_2, \ldots
  \right)
  = \sum_{\lambda \in H^+_N} S_\lambda (y) S_\lambda (x).
  \tag{3.9}
$$
This last formula was stated in \cite{AFMO}.

\head 4. On complete reducibility of certain
$\widehat{gl}\/$-modules
\endhead

Recall that labels of the highest weight $\Lambda =
(\Lambda_j)_{j \in \BZ}$ of the $\widehat\gl\/$-module $L
(\Lambda, c)$ are the numbers $n_i = \Lambda_i - \Lambda_{i + 1}
+ \delta_{i, 0} c\/$ ($i \in \BZ$).  It is clear that a sequence
$\{ n_i\}_{i \in \BZ}$ is the sequence of labels of a highest
weight $\Lambda$ of a $\widehat\gl\/$-module $L (\Lambda, -N)$
that occurs in the decomposition \thetag{3.7} iff the following
properties hold:
$$
\gather
n_i \in \BZ_+ \quad \text{if} \quad i \neq 0 \quad \text{and}
\quad \sum_i n_i = -N, \tag{4.1a}\\
\text{if } n_i \neq 0 \quad \text{and} \quad n_j \neq 0,
\quad \text{then} \quad |i - j| \leq N. \tag{4.1b}
\endgather
$$

We denote, as usual, by $O_{-N}\/$ the category of
$\widehat\gl\/$-modules for which the subalgebra $\fb\/$ acts
locally finitely and $c = -N I$.  All irreducible subquotients of
a module from the category $O_{-N}\/$ are the modules $L
(\Lambda, -N)$.  The following is the main result of this
section.

\proclaim{Theorem 4.1}
If all irreducible subquotients of a $\widehat\gl\/$-module from
the category $O_{-N}$ have labels satisfying conditions
\thetag{4.1a,b}, then this module is a direct sum of irreducible
$\widehat\gl$-modules.
\endproclaim

First, we translate the problem to that for the Lie algebra
$\gl_{fin}$ of matrices $(a_{ij})_{i, j \in \BZ}$ with only a
finite number of non-zero $a_{ij}$'s.  We denote by $\fb_{fin}$ the
subalgebra of upper triangular matrices and by $L (\Lambda)$ the
irreducible $\gl_{fin}$-module defined by the property that there
exists an eigenvector $v_\Lambda$ for $\fb\/$ such that $E_{ii}
v_\Lambda = \Lambda_i v_\Lambda$; we let $\Lambda =
(\Lambda_i)_{i \in \BZ}$.

Consider the homomorphism $\varphi : \gl_{fin} \to \widehat\gl$
defined by:
$$
\varphi (E_{ij}) = E_{ij} \text{ if } i \neq j \text{ or } i = j
\leq 0, \quad
\varphi (E_{ii}) = E_{ii} - K \text{ if } i > 0.
$$
Define a sequence $\widetilde N = (\widetilde N_i)_{i \in \BZ}$
by
$$
\widetilde N_i = 0 \text{ if } i \leq 0, \quad
\widetilde N_i = N \text{ if } i > 0.
$$

Then the set of highest weights that occurs in the decomposition
\thetag{3.7}, when restricted to $\gl_{fin}\/$ via $\varphi$, is
characterised by the following properties:
$$
\gather
\Lambda_i \in \BZ,\;
\Lambda_i = 0 \text{ for } i \leq - N - 1 \text{ and } \Lambda_i
= N \text{ for } i \geq N, \tag{4.2a} \\
\Lambda_i \geq \Lambda_{i + 1} \text{ if } i \neq 0, \tag{4.2b}
\\
\text{if } \Lambda_i \neq \widetilde N_i \text{ and } \Lambda_j
\neq \widetilde N_j, \text{ then } |i - j| \leq N - 1. \tag{4.2c}
\endgather
$$

It is clear that Theorem~4.1 follows from the analogous statement
for the Lie algebra $\gl_{fin}$:

\proclaim{Proposition 4.1}
If $\fb_{fin}$ acts locally finitely in a $\gl_{fin}$-module and
all the irreducible subquotients of this module have highest
weights satisfying conditions
\linebreak
\thetag{4.2a--c}, then this module
is a direct sum of irreducible $\gl_{fin}$-modules.
\endproclaim

Let $S_\infty$ be the group of all permutations $\sigma\/$ of
$\BZ$ such that $\sigma (i) = i\/$ for all but finitely many $i$.
Define the weight $\rho\/$ by $\rho_i = -i$ ($i \in \BZ$).  Since
$\gl_{fin}\/$ is the inductive limit of the Lie algebras $\gl_n$,
by highest weight representation theory for $\gl_n$ (see e.g.\
\cite D) Proposition~4.1 follows from

\proclaim{Proposition 4.2}
Let both $\Lambda = (\Lambda_i)_{i \in \BZ}$ and $M = (M_i)_{i
  \in \BZ}$ satisfy conditions \rom{(4.2a--c)} and let
$$
\sigma (\Lambda + \rho) = M + \rho \text{ for some } \sigma \in
S_\infty.
\tag{4.3}
$$
Then $\Lambda = M$.
\endproclaim

\demo{Proof}
Suppose that $j < - N\/$ is not fixed by $\sigma\/$ and let $j' =
\sigma (j)$.  Then we have:
$$
(\Lambda + \rho)_j = (M + \rho)_j = -j \text{ and } (M +
\rho)_{j'} = -j.
$$
It follows that $j' \in [-N, N - 1]$ and that the transposition
($j j'$) of $j\/$ and $j'\/$ does not change $M + \rho$.  Hence
the permutation $\sigma' = (j j') \sigma\/$ still satisfies
\thetag{4.3} and has fewer non-fixed points outside $[-N, N -
1]$.  We argue similarly in the case $j \geq N$.  Thus we may
assume that
$$
\sigma (j) = j \text{ for } j \notin [-N, N - 1].
\tag{4.4}
$$

Given $a \in \BZ$ let
$$
\Cal J_a = \{ i \in \BZ | a < i \leq N \} \cup
\{ i \in \BZ | - N - 1 < i \leq a - N \}.
$$
By \thetag{4.2c} there exists $b \in [-N, N - 1]$ such that
$$
\Lambda_i = \widetilde N_i \text{ for } i > b \text{ and for } i
\leq b - N.
$$
Then the set $\{ (\Lambda + \rho)_i | i \in \Cal J_b \}$ is a
permutation of the set $[1, N]$ since $ (\Lambda + \rho)_i = N -
i\/ $ for $i > b\/$ and $(\Lambda + \rho)_i = -i\/$ for $i \leq b
- N$; we denote the corresponding bijective map of this set to
the set of integers $[1, N]$ by $\tau_b$.  Similarly choose $b_1
\in [-N, N - 1]$ such that
$$
M_i = \widetilde N_i \text{ for } i > b_1 \text{ and for } i \leq
b_1 - N,
$$
and define the map $\tau_{b_1} : \{ (M + \rho)_i | i \in \Cal
J_b \} \to [1, N]$.

We may assume that in addition to properties \thetag{4.3} and
\thetag{4.4}, $\sigma\/$ has the property
$$
\sigma (i) = \tau^{-1}_{b_1} \tau_b (i) \text{ for } i \in
\Cal J_b.
\tag{4.5}
$$
Indeed, if \thetag{4.5} does not hold for some $i \in \Cal
J_b$, it means that $\sigma (i) \notin \Cal J_{b_1}$.  Then
we have:
$$
(\Lambda + \rho)_i = \tau_b (i), \quad
(M + \rho)_{\tau^{-1}_{b_1} \tau_b (i)} = \tau_b (i) =
(M + \rho)_{\sigma (i)}.
$$
Hence the permutation $\sigma' = \bigl(\sigma (i),
\tau^{-1}_{b_1} \tau_b (i) \bigr) \sigma\/$ still satisfies
\thetag{4.3} and \thetag{4.4} and reduces the number of $i \in
\Cal J_{b}\/$ which do not satisfy \thetag{4.5}.

In order to complete the proof of Proposition~4.2 we need

\proclaim{Lemma 4.1}
Suppose that $\Lambda$, $M$, $\sigma$, $b$ and $b_1$ satisfy
\rom{\thetag{4.2}--\thetag{4.5}}, and assume that $b \geq b_1$.
  Then
  \roster
  \item"(a)" $\sigma (i) = i$ for $b - N < i \leq b_1$

  \item"(b)" $(\Lambda + \rho)_i = N - i \text{ for } b_1 < i
    \leq b$, $(M + \rho)_i = -i \text{ for } b_1 - N < i \leq b
    - N$.
  \endroster
\endproclaim

\demo{Proof}
Consider $b_1 + 1$ numbers $ (\Lambda + \rho)_i\/ $ for $0 \leq i
\leq b_1$.  This is a strictly decreasing sequence of integers
and $ (\Lambda + \rho)_{b_1} \geq N - b_1 $.  None of these
integers is mapped by $\sigma\/$ to $(M + \rho)_j\/$ for $b_1 - N
< j < 0$ since $ (M + \rho)_j < N - b_1 $ for these $j$.  Hence
the strictly decreasing sequence $\bigl\{ ( \Lambda + \rho )_i
\bigr\}_{0 \leq i \leq b_1}$ is mapped by $\sigma\/$ to strictly
decreasing sequence $\bigl\{ (M + \rho )_j \bigr\}_{0 \leq j \leq
  b}$.  So the only possibility for $\sigma\/$ is:
$$
\sigma (i) = i \text{ and } (\Lambda + \rho)_i = (M + \rho)_i
\text{ for } 0 \leq i \leq b_1.
$$
A similar argument works for $b - N < i < 0$, proving (a).

The proof of (b) is similar.  The strictly decreasing sequence
$\bigl\{ (\Lambda + \rho)_i \bigr\}_{b_1 < i \leq b}$ is mapped
by $\sigma\/$ to strictly decreasing sequence $\left\{ (M
+\rho)_j \right\}_{b_1 - N < j \leq b - N}$.  Since $
(\Lambda + \rho)_b \geq N - b\/ $ and $ (M + \rho)_{b_1 - N + 1}
\leq N - b_1 - 1$, the possibility described by (b) is the only
one. \qed
\enddemo

\demo{End of the proof of Proposition \rom{4.2}} Exchanging
$\Lambda\/$ and $M\/$ and replacing $\sigma\/$ by $\sigma^{-1}$,
if necessary, we may assume that $b \geq b_1$.  By \thetag{4.4}
and Lemma~4.1b, $\Lambda_i = M_i\/$ for $i \notin (b - N, b_1]$;
by Lemma~4.1a, $\Lambda_i = M_i\/$ for $i \in (b - N, b_1]$.\qed
\enddemo

\enddemo

\remark{Remark 4.1} Consider the antilinear anti-involution of
the Weyl algebra $W_N$ defined by
$$
(\gamma^i_m)^\dag = \gamma^{* i}_{-m} \quad \text{if} \quad
m < 0, \quad
\gamma^i_m)^\dag = - \gamma^{*i}_{-m} \quad \text{if} \quad m \geq 0.
$$
The unique Hermitean form on $M$, normalized by the condition
that the length of the vacuum vector is 1 and such that the
operator adjoint to $a \in W_N\/$ is $a^\dag$, is positive
definite.  The anti-involution $\dag$ induces the compact
anti-involution on $\gl_N\/$ and the following antilinear
anti-involution on $\widehat\gl$:
$$
E^\dag_{ij} = E_{j i} \quad \text{if} \quad i, j > 0
\quad \text{or}\quad i, j \leq 0, \quad
E^\dag_{i j} = - E_{j i} \quad \text{otherwise},
$$
so that on $\gl_- \oplus \gl_+$ it induces the compact
involution.  Thus all the $\widehat\gl\/$-modules $L (\Lambda,
-N)$ with $\Lambda$ satisfying \thetag{4.1} are unitarizable with
respect to the (non-compact) anti-involution $\dag$.  One may
view \thetag{3.4} as their ``$K\/$-type decomposition.''  Note
also that, in view of \cite{KV} and \cite{O}, these are all
unitarizable highest weight $\widehat\gl\/$- modules with central
charge $-N$.  \endremark

\head 5. A digression on vertex algebras and their twisted
modules \endhead

We explain here the basics of the theory of vertex algebras
(sometimes also called vertex operator algebras) that will be
used in the sequel.  Our definition of a vertex algebra $V\/$
is close to that used in \cite{FKRW} (we do not assume here that
$V\/$ is $\BZ_+$-graded) and one can show ({\it cf.}
\cite{G} and \cite{K2}) that it is equivalent to the original
definition of Borcherds \cite{B1}, \cite{B2}.  The reader may
also consult \cite{FLM} for a different formalism.

Let $V\/$ be a vector space.  A {\it field\/} is a series of the
form
$$
  a (z) = \sum_{n \in \BZ} a_{(n)} z^{- n - 1},
$$
where $a_{(n)} \in \End V$ are such that for each $v \in V\/$ one
has: $a_{(n)} v = 0 \quad \text{for} \quad n \gg 0$.  Here $z\/$
is a formal indeterminate.  We shall often use a different
indexing of the modes of $a (z)$; in such a case the
parenthesis around indices will be dropped, like $a (z) = \sum_n
a_n z^{-n - \Delta}$.

A vertex algebra structure on $V\/$ is a
vector $| 0 \rangle \in V\/$ (called the vacuum vector) and a
linear map of $V\/$ to the space of fields $a \mapsto Y (a, z) =
\sum_{n \in \BZ} a_{(n)} z^{- n - 1}$ (called the state-field
correspondence), satisfying the following axioms:
\roster
\item"(T)" $[T, Y (a, z)] = \partial_z Y (a, z), \text{ where }
  T \in \End V $ is defined by $T (a) = a_{(-2)} | 0
  \rangle$,

\item"(V)" $Y (|0\rangle, z) = I_V, \quad Y (a, z) | 0 \rangle |_{z =
    0} = a$,

\item"(L)" $(z - w)^N [Y (a, z), Y (b, w)] = 0 \quad \text{for}
  \quad N \gg 0$.
\endroster
Here and further the equality means a coefficient-wise equality
of series in $z\/$ or in $z\/$ and $w$.  These axioms are
usually called the translation covariance, the vacuum and the
locality axioms.

Let $\Gamma$ be an additive subgroup of $\BC$ containing
$\BZ$.  A $\Gamma$-twisting of $V\/$ is a $\Gamma /
\BZ$-gradation $V = \bigoplus_{\bar \alpha \in \Gamma / \BZ}
\psup{\bar\alpha} V\/$ such that
$$
  a_{(n)} \psup{\bar \beta} V \subset \psup{\bar \alpha + \bar
    \beta} V \quad \text{if} \quad a \in \psup{\bar \alpha} V.
$$
Here and further $\bar \alpha\/$ stands for the coset $\alpha
+ \BZ$.

Let $M\/$ be a vector space and let $\bar\alpha \in \Gamma / \BZ$.
An $\bar\alpha\/$-twisted $\End M\/$-valued field is a series of
the form $\sum_{n \in \bar\alpha} a^M_{(n)} z^{-n - 1}$ where
$a^M_{(n)} \in \End M\/$ is such that $a^M_{(n)} v = 0$ for $v
\in M$, $n \gg 0$.  A $\Gamma\/$-twisted $V\/$-module $M\/$ is a
linear map from $V\/$ to the linear span of $\Gamma\/$-twisted
$\End M\/$-valued fields, associating to each $a \in \psup{\bar
\alpha} V\/$ an $\bar\alpha\/$-twisted field $Y^M (a, z) =
\sum_{n \in \bar\alpha} a^M_{(n)} z^{-n - 1}$ such that the following
three axioms hold:
\roster
\item"(M1)" $Y^M (1, z) = I_M$,

\item"(M2)" $Y^M (T a, z) = \partial_z Y^M (a, z)$,

\item"(M3)" (twisted Borcherds identity) $\sum^\infty_{j = 0}
  \left(
    \smallmatrix
    m \\
    j
    \endsmallmatrix
  \right) Y^M (a_{(n + j)} b, z) z^{m - j}$
  $$
  = \sum^\infty_{j = 0} (-1)^j
  \left(
    \smallmatrix
    n \\
    j
    \endsmallmatrix
  \right)
  \left(
    a^M_{(m + n - j)} Y^M (b, z) z^j - (-1)^n Y^M (b, z) a^M_{(m
      + j)} z^{n - j}
  \right)
  $$
  for all $m \in \bar\alpha$, $n \in \BZ$.
\endroster

In the case when $\Gamma = \BZ$, $M\/$ is called a
(untwisted) $V\/$-module.  Note that $V\/$ itself is a
$V\/$-module since one can show that properties (M2) and (M3)
with $\bar\alpha = \BZ$ and superscript $M\/$ removed follow from
the axioms (T), (V), and (L) of a vertex algebra (see \cite{K2}).

Letting $n = 0$ in (M3) and substituting $Y^M (b, w) = \sum_{k
  \in \bar\beta} b_{(k)}z^{-k - 1}$, we get
$$
  \left[
    a^M_{(m)}, b^M_{(n)}
  \right] = \sum^\infty_{j = 0}
  \left(
    \smallmatrix
      m \\
      j
    \endsmallmatrix
  \right) (a_{(j)} b)^M_{(m + n - j)}, \quad m \in \bar\alpha,\;
  n \in \bar\beta.
  \tag{5.1}
$$
Letting $m = \alpha \in \bar\alpha$ and $n = -1$ in (M3) we get
$$
  \sum^\infty_{j = 0}
  \left(
    \smallmatrix
      \alpha \\
      j
    \endsmallmatrix
  \right) Y^M (a_{(j - 1)} b, z) z^{-j} =
  : Y^M (a, z) Y^M (b, z):,
  \tag{5.2}
$$
where the normally ordered product is defined, as usual, by
$$
  : Y^M (a, z) Y^M (b, z): = Y^M (a, z)_+ Y^M (b, z) + Y^M (b, z)
  Y^M (a, z)_-
$$
and
$$
  Y^M (a, z)_- = \sum_{j \in \alpha + \BZ_+} a^M_{(j)} z^{-j - 1},
  \quad Y^M (a, z)_+ = Y^M (a, z) - Y^M (a, z)_-
$$
are the {\it annihilation\/} and the {\it creation\/} parts of
$Y^M (a, z)$.  Letting $m = \alpha \in \bar\alpha$ and replacing
$n\/$ by $- n - 1$ with $n \in \BZ_+$ in (M3) gives:
$$
  \sum^\infty_{j = 0}
  \left(
    \smallmatrix
      \alpha \\
      j
    \endsmallmatrix
  \right) Y^M (a_{(-n + j - 1)} b, z) z^{-j} = : \partial^{(n)}
  Y^M (a, z) Y^M (b, z):,
  \tag{5.3}
$$
where $\partial^{(n)}$ stands for $\partial^n_z / n!$.  It is
easy to show that \thetag{5.1} and \thetag{5.2}
along with (M2) imply (M3) (or \thetag{5.1} and \thetag{5.3}
along with (M1) imply (M2) and (M3)).

It is well-known that the space $M\/$ of the metaplectic
representation carries a structure of a vertex algebra with
vacuum vector $| 0 \rangle$ and with the state-field
correspondence defined as follows ($m_i, n_i \in Z_+$; $1 \leq
i_1, \ldots, i_s, j_1, \ldots, j_t \leq N$):
$$
  \multline
    Y
    \left(
      \gamma^{i_1}_{-m_1 - 1} \cdots \gamma^{i_s}_{-m_s - 1}
      \gamma^{* j_1}_{-n_1} \cdots \gamma^{* j_t}_{-n_t} | 0
      \rangle, z
    \right)
    \\
    = : \partial^{(m_1)} \gamma^{i_1} (z) \cdots
    \partial^{(m_s)} \gamma^{i_s} (z) \partial^{(n_1)}
    \gamma^{* j_1} (z) \cdots \partial^{(n_t)}
    \gamma^{* j_t} (z):.
  \endmultline
$$
Here the normally ordered product of more than two fields is
taken from right to left.

Fix a non-zero complex number $s\/$ and let $\Gamma_s\/$
denote the additive subgroup of $\BC$ generated by $s\/$ and
1.  Define a $\Gamma_s\/$-twisting of $M\/$ by letting the
twist of
$$
  \gamma^{i_1}_{m_1} \cdots \gamma^{i_u}_{m_u}
  \gamma^{* j_1}_{n_1} \cdots \gamma^{* j}_{n_t} | 0 \rangle
  \quad \text{equal} \quad
  s (u - t) + \BZ.
$$
We construct a $\Gamma_s\/$-twisted $M\/$-module $M_s\/$ as
follows ({\it cf.} \cite{ABMNF}).  As a vector space we take
$M_s = M\/$ and let
$$
  \gamma^i_s (z) \equiv
  Y^{M_s}
  \left(
    \gamma^i_{-1} | 0 \rangle, z
  \right) : = z^{- s} \gamma^i (z), \quad
  \gamma^{* i}_s (z) \equiv
  Y^{M_s}
  \left(
    \gamma^{* i}_0 | 0 \rangle, z
  \right) : = z^s \gamma^{* i} (z).
$$
By making use of (M1) and \thetag{5.3}, this extends
inductively to a $\Gamma_s\/$-twisted module structure on
$M_s$.  We shall write $Y_s (a, z) = Y^{M_s} (a, z)$ to
simplify the notation.

Denote by $W_{1 + \infty, - N}\/$ the vertex subalgebra of the
vertex algebra $M\/$ generated by the fields
$$
  J^k (z) = \sum^N_{i = 1} : \gamma^i (z) \partial^k_z \gamma^{*
    i} (z):
$$
({\it i.e.}, $W_{1 + \infty, - N}\/$ is the subspace of
$M\/$ consisting of polynomials of the modes of the $J^k (z)$
applied to $| 0 \rangle$).  Note that we have in the domain
$|\epsilon| < |z|$:
$$
  \sum^\infty_{k = 0} J^k (z) \epsilon^k / k! = E (z, z +
  \epsilon).
$$

Due to Corollary~3.1, it follows that
$$
  W_{1 + \infty, - N} = M^{\gl_N}.
  \tag{5.4}
$$
Note that for all $\Gamma_s\/$-twistings of $M\/$ we
have
$$
  W_{1 + \infty, - N} \subset \psup 0 M.
$$
It follows that the restriction of the $\Gamma_s\/$-twisted
$M\/$-module $M_s\/$ to the vertex subalgebra $W_{1 +
  \infty, - N}$ is a (untwisted) $W_{1 + \infty, - N}$-module.
We have by \thetag{5.2} for $1 \leq i \leq N$, $k \in
\BZ_+$:
$$
  : Y_s
  \left(
    \gamma^i_{-1} | 0 \rangle, z
  \right) Y_s
  \left(
    \gamma^{* i}_{-k} | 0 \rangle, z
  \right) : = Y_s
  \left(
    \gamma^i_{-1} \gamma^{*i}_{-k} | 0 \rangle, z
  \right) +
  \left(
    \smallmatrix
      s \\
      k + 1
    \endsmallmatrix
  \right) z^{-k - 1} I.
  \tag{5.5}
$$
Noting that
$$
  \frac{1}{k!}
  J^k (z) = Y
  \left(
    \sum^N_{i = 1} \gamma^i_{-1} \gamma^{* i}_{-k} | 0 \rangle, z
  \right)
  \tag{5.6}
$$
we obtain from \thetag{5.5} and \thetag{5.3}
$$
  \split
    J^k_s (z) & \equiv k! Y_s
    \left(
      \sum^N_{i = 1} \gamma^i_{-1} \gamma^{* i}_{-k} | 0 \rangle, z
    \right) = \sum^N_{i = 1} : \gamma^i_s (z) \partial^k_z \gamma^{*
      i}_s (z) : \\
    & \qquad +
    N \frac{s (s - 1) \cdots (s - k)}{k + 1} z^{-k - 1} I.
  \endsplit
  \tag{5.7}
$$

\head 6. The vertex algebra $W_{1 + \infty, c}$ and its modules
at $c = -N$
\endhead

Let $\CD\/$ be the Lie algebra of complex  regular differential
operators on $\BC^\times$ with the usual bracket, in an
indeterminate $t$.  The elements
$$
  J^l_k = - t^{l + k} (\partial_t)^l \quad
  (k \in \BZ, l \in \BZ_+)
  \tag{6.1}
$$
form a basis of $\CD$.  The Lie algebra $\CD\/$ has the
following 2-cocycle with values in $\BC$ \cite{KP, p.~3310}:
$$
  \Psi (f (t) \partial_t^m, g (t) \partial_t^n) =
  \frac{m! n!}{(m + n + 1)!}
  \Res_{t = 0} f^{(n + 1)} (t) g^{(m)} (t) dt,
  \tag{6.2}
$$
where $f^{(m)} (t) = \partial^m_t f (t)$.  We denote by
$\widehat\CD = \CD \oplus \BC C$, where $C\/$ is the central
element, the corresponding central extension of the Lie algebra
$\CD$.

Another important basis of $\CD\/$ is
$$
  L^l_k = - t^k D^l \quad
  (k \in \BZ, l \in \BZ_+)
  \tag{6.3}
$$
where $D = t \partial_t$.  These two bases are related by the
formula \cite{KR}:
$$
  J^l_k = - t^k D (D - 1) \cdots (D - l + 1).
  \tag{6.4}
$$

Given a sequence of complex numbers $\lambda = (\lambda_j)_{j
  \in \BZ_+}$ and a complex number $c\/$ there exists a unique
irreducible $\widehat\CD\/$-module $L (\lambda, c ;
\widehat\CD)$ which admits a non-zero vector $v_\lambda\/$
such that:
$$
  L^j_k v_\lambda = 0 \quad \text{for} \quad
  k > 0,\; L^j_0 v_\lambda = \lambda_j v_\lambda, \; C = c I.
$$
This is called a highest weight module over $\widehat\CD$ with
highest weight $\lambda\/$ and central charge $c$.  The module $L
(\lambda, c; \widehat \CD)$ is called {\it quasifinite\/} if all
eigenspaces of $D$ are finite-dimensional (note
that $D\/$ is diagonalizable).  It was proved in \cite{KR,
  Theorem 4.2} that $L (\lambda, c; \hat D)$ is a quasi-finite
module if and only if the generating series
$$
\Delta_\lambda (x) = \sum^\infty_{n = 0} \frac{x^n}{n!} \lambda_n
$$
has the form
$$
  \Delta_\lambda (x) = \frac{\phi (x)}{e^x - 1},
  \tag{6.5}
$$
where
$$
  \phi (x) + c = \sum_i p_i (x) e^{r_i x} \quad
  \text{(a finite sum)},
  \tag{6.6}
$$
$p_i (x)$ are non-zero polynomials in $x\/$ such that $\sum_i p_i
(0) = c$ and $r_i\/$ are distinct complex numbers.  The numbers
$r_i\/$ are called the {\it exponents\/} of this module and the
polynomials $p_i (x)$ are called their {\it multiplicities}.  One
has the following nice property of quasi-finite
$\hat\CD\/$-modules:

\proclaim{Proposition 6.1} \cite{KR, Theorem 4.8}.  We have:
$$
  L (\lambda, c; \hat\CD) \otimes L (\lambda', c; \hat\CD) \simeq
  L (\lambda + \lambda', c + c'; \hat\CD)
$$
provided that the difference of any exponent of the first module
and any exponent of the second module is not an integer.
\endproclaim

We call a quasifinite $\widehat\CD\/$-module $L (\lambda, c;
\hat\CD)$ {\it primitive\/} if all multiplicities of its
exponents are integers.  (Note that we required in
\cite{FKRW} the $n_i\/$ to be positive, but we have to drop
this condition here.)

Given $s \in \BC$, we may consider the natural action on the
algebra $\CD$ on the $t^s \BC [t, t^{-1}]$.  Choosing the basis
$v_j = t^{-j + s}$ ($j \in \BZ$) of this space gives us a
homomorphism of Lie algebras $\phi_s : \CD \to \widetilde\gl$:
$$
\phi_s \left(t^k f (D) \right) = \sum_{j \in \BZ} f (-j + s) E_{j
  - k, j}
\tag{6.7}
$$

The homomorphism $\phi_s\/$ lifts to the homomorphism
$\hat\phi_s\/$ of the corresponding central extensions as follows
\cite{KR}:
$$
\align
\hat\phi_s
\left(
  L^j_k
\right) & = \phi_s
\left(
  L^j_k
\right) \text{ if } k \neq 0, \\
\hat\phi_s
\left(
  e^{x D}
\right) & = \phi_s
\left(
  e^{x D}
\right) - \frac{e^{s x} - 1}{e^x - 1} K, \quad
\hat\phi_s (C) = K.
\endalign
$$
It is straightforward to check using \thetag{6.4} that in the
basis $J^n_k\/$ this homomorphism becomes:
$$
\aligned
\hat\phi_s
\left(
  J^n_k
\right) & = \phi_s
\left(
  J^n_k
\right) \text{ if } k \neq 0, \\
\hat\phi_s
\left(
  J^n_0
\right) & = \phi_s
\left(
  J^n_0
\right) - \frac{s (s - 1) \cdots (s - n)}{n + 1} K, \quad
\hat\phi_s (C) = K.
\endaligned
\tag{6.8}
$$

The following proposition is a special case of Theorem~4.7 and
formula \thetag{5.6.5} from \cite{KR}.

\proclaim{Proposition 6.2}
Let $L (\Lambda, c)$ be the irreducible $\widehat\gl\/$-module
with the highest weight $\Lambda = (\Lambda_j)_{j \in \BZ}\/$ and
central charge $k$.  Let $n_j = \Lambda_j - \Lambda_{j + 1} +
\delta_{j, 0} c\/$ be the labels of $\Lambda$.  Then when
restricted to $\hat\phi_s (\hat\CD)$ the module $L (\Lambda, c)$
becomes an irreducible quasifinite $\hat\CD\/$-module with
exponents $s - j\/$ ($j \in \BZ$) of multiplicity $n_j\/$ (and
central charge $c$).
\endproclaim

Recall now that the $\hat\CD\/$-module $L(0, c; \hat\CD)$ has a
canonical structure of a vertex algebra with the vacuum vector
$| 0 \rangle = v_0$ and generated by the fields $J^k (z) =
\sum_{m \in \BZ} J^k_m z^{-m - k - 1}$ \cite{FKRW}.  The
following proposition is immediate from \thetag{5.4},
\thetag{5.7} and \thetag{6.8}.

\proclaim{Proposition 6.3}
We have an isomorphism of vertex algebras:
$$
L (0, -N; \hat\CD) \simeq W_{1 + \infty, -N}
$$
under which the fields (denoted by the same symbol) $J^k (z)$
correspond to each other.
\endproclaim

A $L (0, c; \hat\CD)$-module $M\/$ is called a positive energy
module if the operator $J^{1M}_0$ is diagonalizable and the set
of real parts of its eigenvalues is bounded below.  It is clear
that the irreducible positive energy $L (0, c; \hat\CD)$-modules
are some of the modules $L (\lambda, c; \hat\CD)$, and the
problem is which of the $\Delta_\lambda (x)$ may occur.  As was
pointed out in \cite{FKRW}, all of them occur if $c \notin \BZ$.
One of the main results of \cite{FKRW} is that the irreducible
positive energy $L (0, c; \hat\CD)$-modules with $c \in \BZ_+$
are precisely the primitive modules with non-negative
multiplicities of exponents.

We address now this problem in the remaining case $c = -N$, where
$N\/$ is a positive integer.

\proclaim{Theorem 6.1}
Viewed as a $W_{1 + \infty, -N\/}$-module the module $M_s\/$
decomposes in a direct sum (with multiplicities)
of all primitive modules for which the
set of exponents lies in $s + \BZ$, the difference of any two
exponents does not exceed $N\/$ and the multiplicity of only one
exponent is negative.
\endproclaim

\demo{Proof}
The proof follows from remarks made at the end of Section~4, Theorem~3.1,
and Proposition~6.2.\qed
\enddemo

Theorem~6.1 along with Proposition~6.1 imply

\proclaim{Theorem 6.2}
Let $R\/$ be the set of exponents of a primitive
$\hat\CD\/$-module $V\/$ with central charge $-N$.  Let $s_1,
s_2, \ldots$ be all exponents with negative multiplicity.  Assume
that
\roster
\item"(i)" $s_i - s_j \notin \BZ$ if $i \neq j$.
\endroster

Let $R_i = \{r \in R | r - s_i \in \BZ \}$.  Assume that

\roster
\item"(ii)" $R = \bigcup_i R_i$.
\endroster

Let $N_i = -$ (sum of multiplicities of all exponents from
$R_i$).  Assume that

\roster
\item"(iii)" difference of any two exponents from $R_i$ does
  not exceed $N_i$.
\endroster

Then $V$ is a $W_{1 + \infty, -N}$-module.
\endproclaim

\proclaim{Conjecture 6.1}
Theorem\/~\rom{6.2} lists all irreducible $W_{1 + \infty,
  -N}$-modules.
\endproclaim

\head 7. Charged free fermions \endhead

In this section we show how to recover the main results of the
paper \cite{FKRW} (and to obtain some new results along the way)
using the method of this paper applied to $N\/$ pairs of free
charged fermionic fields
$$
\psi^i (z) = \sum_{m \in \BZ} \psi^i_m z^{-m -1} \quad
\text{and} \quad
\psi^{*i} (z) = \sum_{m \in \BZ} \psi^{*i}_m z^{-m}.
$$
Recall that this is a collection of local odd fields with the OPE
\thetag{2.1} and the vacuum vector $| 0 \rangle$ subject to conditions
\thetag{2.2}.  Here and further we substitute $\psi\/$ in place
of $\gamma$.  In other words, we have a unital associative
algebra $C_N$, called the Clifford algebra, on generators
$\psi^i_m$, $\psi^{*i}_m\/$ ($i = 1, \ldots, N$; $m \in \BZ$)
with the following defining relations ({\it cf.} \thetag{2.3}):

$$
\left[
  \psi^{*i}_m, \psi^j_n
\right]_+ \equiv \psi^{*i}_m \psi^j_n + \psi^j_n \psi^{*i}_m =
\delta_{i j} \delta_{m, -n},
$$
and all other anticommutators equal zero.  The algebra $C_N\/$
has a unique irreducible representation in a vector space $F$,
called the spin representation, such that there exists a non-zero
vector $|0 \rangle$ satisfying \thetag{2.2}.  An important
difference with the bosonic case is that $F\/$ is a superspace
with the parity
$$
p (| 0 \rangle) = \bar 0, \quad
p
\left(
  \psi^i_m
\right) = p
\left(
  \psi^{*i}_m
\right) = \bar 1.
$$

We introduce the fields $e^{ij} (z)$ and $E (z, w)$ by the same
formulas as in Section~2.  Then we have to change the sign in the
last summand of \thetag{2.4}, meaning that the $e^{ij}_m\/$ form
a representation in $F\/$ of the affine Kac-Moody algebra of
level 1.  Formulas \thetag{2.5} and \thetag{2.6} remain
unchanged.  We have to change the sign in the last summand of
\thetag{2.7}, meaning that the operators $E_{ij}\/$ form a
representation in $F\/$ of the Lie algebra $\widehat \gl\/$ with
central charge $N$.

In the same way as in Section~2, we prove

\proclaim{Proposition 7.1}
\rom{(a)} The associative algebra $(C_N)^{\gl_N}$ is generated
by the elements $E_{ij}$ ($i, j \in \BZ$).

\rom{(b)} As a $\gl_N$-module, $F$ is isomoprhic to the exterior
algebra over the $\gl_N$-module $U^*_- \oplus U_+$, hence, in
particular, decomposes into a direct sum of irreducible
finite-dimensional $\gl_N$-modules.

\rom{(c)} Each $\gl_N$ isotypic component $F_E$ is an irreducible
$\gl_N \oplus \widehat\gl$-module.
\endproclaim

Denote by $H\/$ the set of sequences $(\lambda_i)_{i \in \BZ}$
such that $ (\lambda_i)_{i > 0} \in H_+ $ and $ (\lambda_i)_{i
  \leq 0} \in H_- $.  Define a map $T : H \to H\/$ by:
$$
\left(
  \lambda^T
\right)_i = \# \{ j | \lambda_j \geq i \} \text{ for } i
> 0,\qquad
\left(
  \lambda^T
\right)_i = - \# \{ j | \lambda_j \leq i - 1 \} \text{ for } i
\leq 0.
$$
In other words, when restricted to $H_-$ and to $H_+$, $T\/$ is
the usual transposition of the Young diagram with respect to the
main diagonal.

\proclaim{Lemma 7.1}
The $\gl_N \oplus \gl_\pm$-module $\Lambda
\left(
  \BC^{\pm N} \otimes \BC^{\pm \infty}
\right)$ has the following decomposition into a direct sum of
irreducible modules:
$$
\Lambda
\left(
  \BC^{\pm N} \otimes \BC^{\pm \infty}
\right) = \bigoplus_{\lambda \in H^\pm_N}
\left(
  L_N (\lambda) \otimes L_\pm (\lambda^T)
\right).
$$
\endproclaim

\demo{Proof}
The proof follows from the other of Cauchey's formulae \cite{M}:
$$
\prod^N_{i = 1} \prod^\infty_{j = 1} (1 + x_i y_j) =
\sum_{\lambda\in H^+_N} \ch L_N (\lambda) \ch L_+ (\lambda^T).
\eqno\qed
$$
\enddemo

In the same way as in Section~2 we derive now the following
result.

\proclaim{Proposition 7.2}
The following is a decomposition of $F$ as a $\fg$-module:
$$
F = \bigoplus_{\Sb \lambda \in H^-_N \\
  \mu \in H^+_N \\
  \nu \in H_N
  \endSb} c^\nu_{\lambda, \mu} L_N (\nu) \otimes L_- (\lambda^T)
\otimes L_+ (\mu^T).
$$
\endproclaim

We are in a position now to prove the main result of this
section.

\proclaim{Theorem 7.1}
With respect to $\gl_N \oplus \widehat \gl$, the spin
representation $F$ decomposes as follows:
$$
F = \bigoplus_{\nu \in H_N} L_N (\nu) \otimes L
\left(
  \Lambda (\nu)^T, N
\right),
\tag{7.1}
$$
where $\Lambda (\nu)$ is defined by \thetag{3.5}.
\endproclaim

It is clear that the proof of this theorem reduces to the
following lemma.

\proclaim{Lemma 7.2}
For any $ (\lambda; \mu) \in S_\nu$ (where $S_\nu$ is defined in
Section~3) we have
$$
\Lambda (\nu)^T \geq
\left(
  \lambda^T; \mu^T
\right).
$$
\endproclaim

\demo{Proof}
Define the following operations on the set of pairs $ (\lambda;
\mu) \in H^-_N \times H^+_N$:
$$
\align
T^{(1)}_{i j} : & \lambda_i \to \lambda_i - 1, \quad \lambda_j
\to \lambda_j + 1 \text{ for } i < j, \\
T^{(2)}_{ij} : & \quad \mu_i \to \mu_i - 1, \quad \mu_j \to \mu_j
+ 1 \text{ for } i < j, \\
T^{(3)}_{ij} : & \quad \lambda_i \to \lambda_i + 1, \quad
\mu_j \to \mu_j - 1 \text{ for any } i, j.
\endalign
$$
We will say that $ (\lambda'; \mu') \succ (\lambda; \mu) $ if
one gets $ (\lambda'; \mu') $ from $ (\lambda; \mu) $ by a
sequence of operations $T^{(a)}_{i j}$, $a = 1, 2, 3$.

It is straightforward to show
$$
(\lambda'; \mu') \succ (\lambda; \mu) \text{ iff }
\left(
  {\lambda'}^T; {\mu'}^T
\right) >
\left(
  \lambda^T; \mu^T
\right),
\tag{7.2}
$$
where $>$ is the usual partial order on the set of weights of
$\widehat\gl$.

Fix $\nu \in H_N\/$ and define $p\/$ as in \thetag{3.5}.  Let $r
\in [1, p]$, $s \in [p+1, N]$.  Recall that by Lemma~3.3 we have:
$$
\mu_r \geq \nu_r, \quad \lambda_s \leq \nu_s.
$$
Applying $T^{(1)}_{rs}\/$ for $s\/$ such that $\lambda_s <
\nu_s\/$ and any $r\/$ sufficiently many times to $ (\lambda;
\mu) $ we shall get $ (\lambda'; \mu') $ such that $\lambda'_s =
\nu_s\/$ for each $s$.  Similarly, applying $T^{(2)}_{rs}$ for
$r\/$ such that $\mu_r > \nu_r$, we shall get $ (\lambda'';
\mu'') $ such that $\lambda_s'' = \nu_s''$ and $\mu_r'' =
\nu_r''$ for each $r\/$ and $s$.  It is clear that $q_r'' \geq 0$
and $q_s'' \leq 0$ and $\sum_i q_i'' = 0$ for $ (\lambda'';
\mu'') $.  Applying $T^{(3)}_{r, s}$ for non-zero $q_r\/$ and
$q_s\/$ we get $ (\tilde\lambda; \tilde \mu) $ with the $\tilde
q_i = 0$.  It is clear that $(\tilde\lambda; \tilde \mu) \in
S_\nu\/$ and $(\tilde\lambda; \tilde \mu) \succ (\lambda; \mu) $.
Thus, we obtain that $\Lambda (\nu) \succ (\lambda; \mu) $.  This
together with \thetag{7.2} proves Lemma~7.2.\qed
\enddemo

\remark{Remark \rom{7.1}}
It is easy to see that decomposition \thetag{7.1} coincides with
the decomposition \cite{FKRW, \thetag{6.3}}.
\endremark

The first consequence of Theorem~7.1 is one of the main results
of \cite{FKRW}:
$$
F^{\gl_N} \simeq L (0, N) \text{ as } \widehat\gl\text{-modules}.
\tag{7.3}
$$
As in Section~3, another corollary of Theorem~7.1 and
Proposition~7.2 is a character formula:
$$
\tr_{L (\Lambda, N)} \diag
\left(
  \ldots, y^{-1}_2, y^{-1}_1; x_1, x_2, \ldots
\right) =
\sum_{\lambda \in H^+_N} c^\nu_{\lambda^* \mu} S_{\lambda^T} (y)
S_{\mu^T} (x),
\tag{7.4}
$$
a special case of which is the following formula stated in
\cite{AFMO}:
$$
\tr_{L (0, N)} \diag
\left(
  \ldots, y^{-1}_2, y^{-1}_1; x_1, x_2, \ldots
\right) =
\sum_{\lambda \in H^+_N} S_{\lambda^T} (y) S_{\lambda^T} (x).
\tag{7.5}
$$
These formulas look quite different from \cite{FKRW,
  \thetag{2.7}}.

\remark{Remark \rom{7.2}}
The $\widehat\gl$-modules that occur in \rom{\thetag{7.1}} are
precisely all the modules with central charge $N$ and dominant
integral highest weights $\Lambda$ (\rom{i.e.}, $\Lambda_i \in
\BZ$ and $\Lambda_i \geq \Lambda_j$ if $i \leq j$).  Hence we
have a complete reducibility theorem analogous to Theorem~4.1.
\endremark

\remark{Remark \rom{7.3}}
Consider the antilinear anti-involution of the algebra $C_N$
defined by
$$
\left(
  \psi^i_m
\right)^\dag = \psi^{* i}_{-m}.
$$
The unique Hermitean form on $F$ defined as in Remark~\rom{4.1}
is positive definite.  The anti-involution $\dag$ induces the
usual compact anti-involutions ($A \to \psup{t\,}{\bar A}$) on
$\gl_N$ and $\widehat\gl$, and the anti-involution $\sigma$ on
$\widehat \mathD$ \rom(see \cite{KR}\rom).
\endremark

The space $F\/$ of the spin representations carries a canonical
structure of a vertex superalgebra defined in the same way as in
Section~5 for $M$.  The $\Gamma_s\/$-twised $F\/$-modules $F_s\/$
are defined in the same way too.  The fields $J^k (z)$ defined in
the same way as in Section~5, generate a vertex subalgebra of the
vertex algebra $F\/$ denoted by $W_{1 + \infty, N}$.  Due to
\thetag{7.3}, we have:
$$
W_{1 + \infty, N} = L^{\gl_N}.
\tag{7.6}
$$
We also have a formula similar to \thetag{5.5} except for the
change of sign of the last summand.

In the same way as in Section~6, we have an isomorphism of vertex
algebras \cite{FKRW}:
$$
L (0, N; \widehat D) \simeq W_{1 + \infty, N}.
\tag{7.7}
$$

In the same way as in Section~6, we prove the following theorem.
(Remark~5.2 of \cite{FKRW} should be corrected accordingly.)

\proclaim{Theorem \rom{7.2}}
Viewed as a $W_{1 + \infty, N}$-module, the $F$-module $F_s$
decomposes in a direct sum (with multiplicities) of all primitive
modules for which the set of exponents lies in $s + \BZ$ and
multiplicities of all exponents are positive.
\endproclaim

Using Proposition~6.2 and Theorem~7.2, we see that all primitive
modules whose exponents have positive multiplicities are $W_{1 +
  \infty, N}\/$-modules.  It is shown in \cite{FKRW} that these
are all irreducible $W_{1 + \infty, N}\/$-modules.  In
particular, due to Remark~7.2, any positive energy module of the
vertex algebra $W_{1 + \infty, N}$ is completely reducible.  Of
course a similar statement for $W_{1 + \infty, -N}$ follows from
Conjecture~6.1.

\enddocument